\documentclass[12pt]{article}
\usepackage[symbol]{footmisc}

\usepackage[left=2.5cm,top=2.5cm,right=2cm,bottom=2.5cm]{geometry}
\usepackage{amsmath, amssymb}

\usepackage{graphicx,subfigure,epsfig,epsf,color}
\usepackage{slashed}
\usepackage{feynmf}
\usepackage{graphicx}
\usepackage{graphics}
\usepackage{epstopdf}
\usepackage{subfigure}
\usepackage{amsfonts}
\usepackage{sectsty}
\usepackage{hyperref}
\usepackage{cite}
\usepackage{pdflscape}

\begin{document}
 \date{}
%%%%%%%
  \title{Inflationary cosmology- A new approach using Non-linear electrodynamics}
 \maketitle
% \date{}
\begin{center}
 Payel Sarkar\footnote{p20170444@goa.bits-pilani.ac.in},~Prasanta Kumar Das\footnote{Corresponding author:~pdas@goa.bits-pilani.ac.in} and Gauranga Charan Samanta\footnote{gauranga@goa.bits-pilani.ac.in} \\
 \end{center}
 \begin{center}
 Birla Institute of Technology and Science-Pilani, K. K. Birla Goa campus, NH-17B, Zuarinagar, Goa-403726, India
 \end{center}
 \vspace*{0.25in}
 \begin{abstract}
 We explore a new kind of field of nonlinear electrodynamics(NLED) which acts as a source of gravity and can accelerate the universe during the inflationary era. We propose a new type of NLED lagrangian which is charecterized by two paremetrs $\alpha$ and $\beta$. We investigate the classical stability and causality aspects of this model by demanding that the speed ($C_s = \frac{dP}{d\rho}$) of the sound wave  $C_s^2 > 0$ and $C_s \le 1$ and find that $0 < C_s^2 < 1$ corresponds to $0.25 \le \alpha \le 0.4$ and $0.6 \le \beta B^2 \le 1$. A study of the deceleration parameter($q = \frac{1}{2}(1 + 3 \omega)$, $\omega = P/\rho$ being the equation of state parameter) suggests that the value $q < 0$ (i.e. $\omega < -1/3$ and $\ddot{a}(t) > 0$ ( the accelerating universe)) requires  $\beta B^2 \ge 0.13$. During inflation, the energy density $\rho_B$ is found to be maximum and is given by $\rho_B^{max} = 0.65/\beta$ corresponding to $\alpha = 0.3$. The magnetic field necessary to trigger the inflation, is found to be $B (= B_{max}) \simeq \sqrt{\frac{0.4 \rho_B^{max}}{0.65}} = 4 \times 10^{51}~{\rm Gauss}$, where $\rho_{B}^{max}( =10^{64}~{\rm GeV}^4 $) is the energy density of the universe during inflation.  The model also predict the e-fold number $N = 71$, which agrees with the experimental result.
 \end{abstract}
 \section{Introduction}
 The discovery of general theory of relativity by Albert Einstein in 1915 enabled us to come up with a compelling and testable theory of the universe. The realization that the universe is expanding with acceleration is confirmed by the observation of Cosmic Microwaves Background(CMB) \cite{KT, CMB} and redshift of Ia type Supernova \cite{Riess, Perlmutter}. To explain this cosmic acceleration of the universe one introduce a cosmological constant $\Lambda$ in the Einstein-Hilbert action with the equation of state $P = \omega \rho $ (with $\omega = -1$ and $P$ and $\rho$ the pressure and the energy density of the universe, respectively). The fluid with this property is called dark energy(DE)\cite{Caroll,PR}. Although, the idea of cosmological constant($\Lambda$) is rather simpler to take into account the cosmic acceleration, however it faces some problems due to mismatch between theory and observation.  A typical dark energy model consists of a  scalar field coupled with gravity which can drive the universe to accelerate\cite{Baumann}. Another way to explain the inflation of the universe is to modify general relativity by introducing $f(R)$ type gravity. As the choice of $f(R)$ is not unique, there exists a host of modified gravity models depending on $f(R)$\cite{Nojiri,Clifton}. 
%  
%  The standard cosmological model based on FRW geometry with Maxwell's electrodynamics as its source has the unavoidable cosmological singularity at the very beginning where the curvature and energy density everything blow up - the model enters into the domain of of non-applicability. A host of models proposes the cosmological solution of the primordial singularity. They are based on different mechanisms, such as cosmological constant\cite{Sitter}, nonminimal coupling\cite{NS}, nonlinear curvature terms\cite{MB}etc. The homogeneous and non-singular isotropic FRW cosmological models can be realized by considering local covariant and gauge invariant lagrangian generalization of Maxwell's electrodynamics (dubbed as nonlinear electrodynamics(NLED))\cite{LKNS} and in cosmological setting these theories have been explored mainly in context of 'magnetic universe' filled with Born-Infeld type nonlinear electromagnetic field\cite{GB}. \\
 At the same time, it is widely believed that the universe in early era was dominated by the electromagnetic field which was very strong and highly nonlinear. Such a non-linear electrodynamics(NLED) coupled with gravity can leads to negative pressure, drive acceleration of the universe and can explain Inflation \cite{Ovgun,Kruglov1,Kruglov2, Kruglov3}. \\
\noindent In this paper, we study a model in which gravity couples with non-linear electrodynamics in which the source of gravity is non-linear electromagnetic field. 
%In general relativity it is found that if the source of gravity is non-linear electromagnetic field, then it can play a profound role in cosmology. 
We investigate the accelerated expansion (i.e. inflation) of the early universe filled up by the magnetic field predominantly. \\
 \noindent The paper is organized as follows. In Sec. 2, we describe the model of nonlinear electrodynamics with  dimensional parameters $\beta$ and $\alpha$ in magnetic universe. We obtain the pressure($P_B$), energy density($\rho_B$), equation of state parameter($\omega$) in terms of the magnetic field $B$ and the parameters $\alpha$, $\beta$. In section 3, we investigate the inflationary expansion in a magnetic universe. We first look at the classical stability and causality aspects of the inflationary expansion in the parameter space $\alpha$ and $\beta$ and analyze the behaviour of several variables of inflationary expansion, $B(t)$, Hubble parameter $H(z)$, effective potential $V_{eff}$, $P_B$, $\rho_B$ and $\omega$. We made an estimate of the magnetic field of the inflationary phase. We predict the e-fold number($N$) produced during  this inflation by the nonlinear $B$-field. In Sec. 4. we summarize and then conclude. 
%  We investigate the classical stability and causality aspects of this model of inflationary expansion by demanding that the speed ($C_s = \frac{dP}{d\rho}$) of the sound wave  $C_s^2 > 0$ and $C_s \le 1$ and find that $0 < C_s^2 < 1$ corresponds to $0.25 \le \alpha \le 0.4$ and $0.6 \le \beta B^2 \le 1$. A study of the deceleration parameter($q = \frac{1}{2}(1 + 3 \omega)$, $\omega = P/\rho$ being the equation of state parameter) suggests that the value $q < 0$ (i.e. $\omega < -1/3$ and $\ddot{a}(t) > 0$ ( the accelerating universe)) requires  $\beta B^2 \ge 0.13$. During inflation, the energy density $\rho_B$ is found to be maximum and is given by $\rho_B^{max} = 0.65/\beta$ corresponding to $\alpha = 0.3$. The magnetic field necessary to trigger the inflation, is found to be $B (= B_{max}) \simeq \sqrt{\frac{0.4 \rho_B^{max}}{0.65}} = 4 \times 10^{51}~{\rm Gauss}$, where $\rho_{B}^{max}( =10^{64}~{\rm GeV}^4 $) is the energy density of the universe during inflation. The magnetic field $B$ is found to decrease with time $t$. The model also predict the e-fold number $N = 71$ (a value close to the experimental finding). 
 
%  The cosmology of the universe in presence of stchastic magnetic field is studied in Sec. 3.  We study the inflation of this magnetic universe in which the acceleration is driven by the magnetic field($B$) and  \\
\noindent In our analysis, we have taken the units $\mu_0=\epsilon_0=c= 8\pi G=1$ and the metric signature $\eta=diag(,-,+,+,+)$.
\section{A model of Non-Linear Electrodynamics}
In the early days of our universe expansion, it is assumed that our universe was filled up with highly non-linear and strong electromagnetic field. Therefore it is quite  natural to expect that the laws of classical electrodynamics\cite{Jackson} gets modified for this non-linear strong electromagnetic field. Born and Infeld  first introduced non linear electrodynamics(NLED) into the gravity theory\cite{BI}. The NLED theory when coupled with gravity can describe the Inflation of early universe. The action of such NLED theory coupled with gravity is given by,
\begin{equation}
 \mathcal{S}=\int d^4x \sqrt{-g} ~\left({\mathcal{L}}_{NED}+\frac{1}{2} R\right) \label{action}
\end{equation}
\noindent where $R$ is the Ricci Scalar and we propose a new form of ${\mathcal{L}}_{NED}$ as follows 
\begin{equation}
 {\mathcal{L}}_{NED} =-\frac{\mathcal{F}e^{-\beta\mathcal{F}}}{(\beta\mathcal{F}+\alpha)^2}\label{Lagrangian}
 \end{equation} 
Here $\beta$ is a constant of dimension $[M]^{-4}$ and $\alpha$ is a dimensionless parameter. In above, $\mathcal{F}=(1/4)F_{\mu\nu}F^{\mu\nu}=(B^2-E^2)/2$ where $F_{\mu\nu}=\partial_{\mu}A_{\nu}-\partial_{\nu}A_{\mu}$is the field strength tensor. Note that in $\beta \rightarrow 0$ and $\alpha\rightarrow 1$ limit, the lagrangian reduces to the usual classical Maxwell's electrodynamics. On varying the action (Eq.~(\ref{action})) we derived the Einstein's equation and NLED field equation as follows,
\begin{equation}
 R_{\mu\nu}-\frac{1}{2}g_{\mu\nu}R=-T_{\mu\nu}
\end{equation}
\begin{equation}
 \partial_{\mu}\left(\sqrt{-g}F^{\mu\nu}\frac{\partial {\mathcal{L}}_{NED}}{\partial\mathcal{F}}\right)=0
\end{equation}
The energy-momentum tensor is given by 
\begin{equation}
 T_{\mu\nu}=\frac{\partial {\mathcal{L}}_{NED}}{{\partial\mathcal{F}}} F_{\mu\alpha}F^{\alpha}_{\nu}- {\mathcal{L}}_{NED} g_{\mu\nu}
\end{equation}\label{tmunu}
The density and the pressure can be derived from Eq.~(\ref{tmunu}) as follows,
\begin{equation}
 \rho_{NED}=-{\mathcal{L}}_{NED}-E^2\frac{\partial\mathcal{L}}{\partial\mathcal{F}}
 =\frac{e^{-\beta\mathcal{F}}}{(\beta\mathcal{F}+\alpha)^2}\left[\mathcal{F}-\frac{2\beta\mathcal{F}E^2}{(\beta\mathcal{F}+\alpha)}+E^2-\beta\mathcal{F}E^2\right]
\end{equation}
and
\begin{equation}
 P={\mathcal{L}}_{NED}+\frac{E^2-2B^2}{3}\frac{\partial\mathcal{L}}{\partial\mathcal{F}}=\frac{e^{-\beta\mathcal{F}}}{(\beta\mathcal{F}+\alpha)^2}\left[-\mathcal{F}+\frac{E^2-2B^2}{3}\left(\frac{2\beta\mathcal{F}}{(\beta\mathcal{F}+\alpha)}-1+\beta\mathcal{F}\right)\right]
\end{equation}
According to the standard cosmological models a symmetry in the direction holds (i.e. the universe is isotropic) and the averaged magnetic field $\langle B \rangle =0$. Again for the magnetic universe, we set the electric field ${\bf{E}} = 0$, as the average electric field $\bf{E}$ is screened by the charged primordial plasma (the state field ${\bf{E}}$
of our early universe) and find the density $\rho(=\rho_B)$ and pressure $p(=p_B)$ of the magnetic fluid as 
\begin{equation}
 \rho_{B}=\frac{2B^2e^{-\beta B^2/2}}{(\beta B^2+\alpha)^2},~
% \end{equation}
% \begin{equation}
 P_{B}=-\frac{4B^2e^{-\beta B^2/2}}{3(\beta B^2+\alpha)^3}\left[\beta^2 B^4+\beta B^2\left(\frac{7}{2}+2\alpha\right)-\alpha\right]
\end{equation}
This gives
\begin{equation}
 \rho_B+3P_B=-\frac{4B^2e^{-\beta B^2/2}}{3(\beta B^2+\alpha)^3}\left[\beta^2B^4+\beta B^2(2\alpha+3)-2\alpha\right]\label{rhoplus3P}
\end{equation}
and
\begin{equation}
 \rho_B+P_B=\frac{2B^2e^{-\beta B^2/2}}{3(\beta B^2+\alpha)^3}\left[\alpha(8-4\beta B^2)-2\beta^2B^4-4\beta B^2\right]\label{rhoplusP}
\end{equation}
%%%%%%
The Raychaudhuri equation is given by 
\begin{equation}
 3\frac{\ddot a}{a}=-\frac{1}{2}\left(\rho_B +3P_B\right)
\end{equation}
Note that we have set $\kappa^2 (= 8 \pi G) = 1$ in above. The acceleration  $\ddot a>0$ requires $\rho_B+3P_B < 0$ (violation of energy condition) and this gives,
\begin{equation}
 \beta B^2>-\frac{1}{2}(2\alpha+3)+\frac{1}{2}\sqrt{(4\alpha^2+20\alpha+9)}
\end{equation}
From the conservation of energy-momentum tensor($ \nabla_{\mu} T^{\mu\nu} =0$), we find the equation of continuity
\begin{equation}
 \dot{\rho_B}+3\frac{\dot a}{a}(\rho_B+P_B)=0 \label{eqnofcontinuity}
\end{equation}
which gives,
\begin{equation}
 \frac{2Be^{-\beta B^2/2}}{3(\beta B^2+\alpha)^3}(\dot{B}+2HB)(4\alpha-2\beta B^2-\beta^2B^4-2\alpha\beta \dot{B}^2)=0
\end{equation}
where $H=\frac{\dot a(t)}{a(t)}$ is the Hubble cosntant. Integrating this, we find $B(t)$ evolves with $a(t)$ as 
\[B(t)=\frac{B_0}{a^2(t)}\label{BB}\]
where $B_0$ ($B(t)$ at $t=t_0$) is the present value of magnetic field. Using this, we can rewrite $\rho_B$ and $P_B$ as
\begin{equation}
 \rho_B=\frac{2B_0^2}{a^4}e^{-\beta B_0^2/2a^4}\left(\frac{\beta B_0^2}{a^4}+2\alpha\right)^{-2}\label{density}
\end{equation}
\begin{equation}
P_B=-\frac{4B_0^2}{3a^4}\frac{e^{-\beta B_0^2/2a^4}}{\left(\frac{\beta B_0^2}{a^4}+2\alpha\right)^3}\left[\frac{\beta^2B_0^4}{a^8}+\frac{\beta B_0^2}{a^4}\left(\frac{7}{2}+2\alpha\right)-\alpha\right]\label{pressure}
\end{equation}
From the equation of state(EoS) $P_B = \omega {\rho_B}$, we find 
\begin{equation}
 \omega=\frac{P_{B}}{\rho_B}=-\frac{2}{3}\left(\frac{\beta B_0^2}{a^4}+2\alpha\right)^{-1}\left[\frac{\beta^2B_0^4}{a^8}+\frac{\beta B_0^2}{a^4}\left(\frac{7}{2}+2\alpha\right)-\alpha\right]\label{omega}
\end{equation}

\noindent Note that $a\rightarrow \infty$, $\omega=\frac{1}{3}$(Radiation dominated universe).
\section{Inflationary expansion: causality and classical stability}
\begin{figure}[!h]
 \centering
  \includegraphics[width=8.0cm]{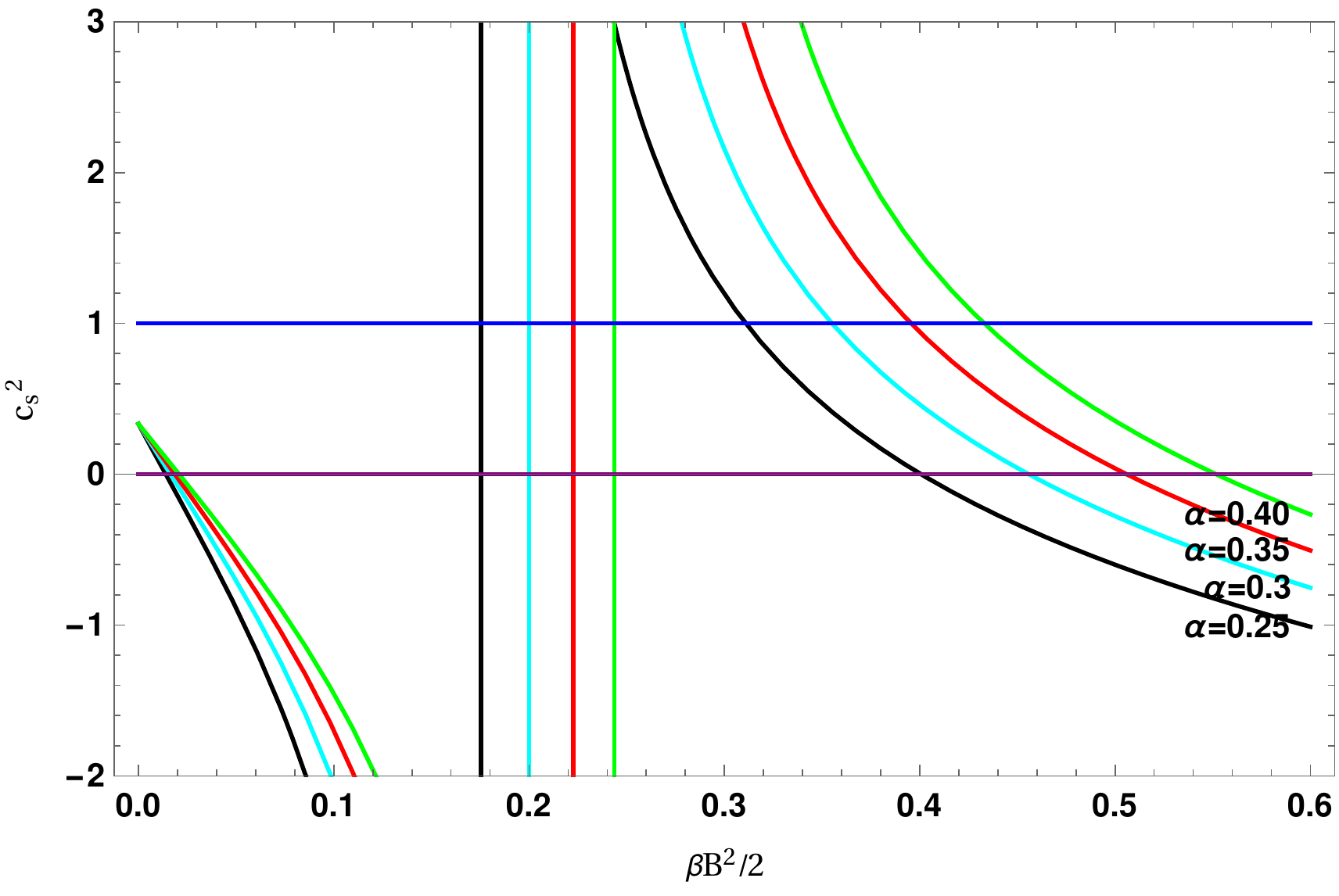}\hspace*{0.1in}
  \includegraphics[width=6.5cm]{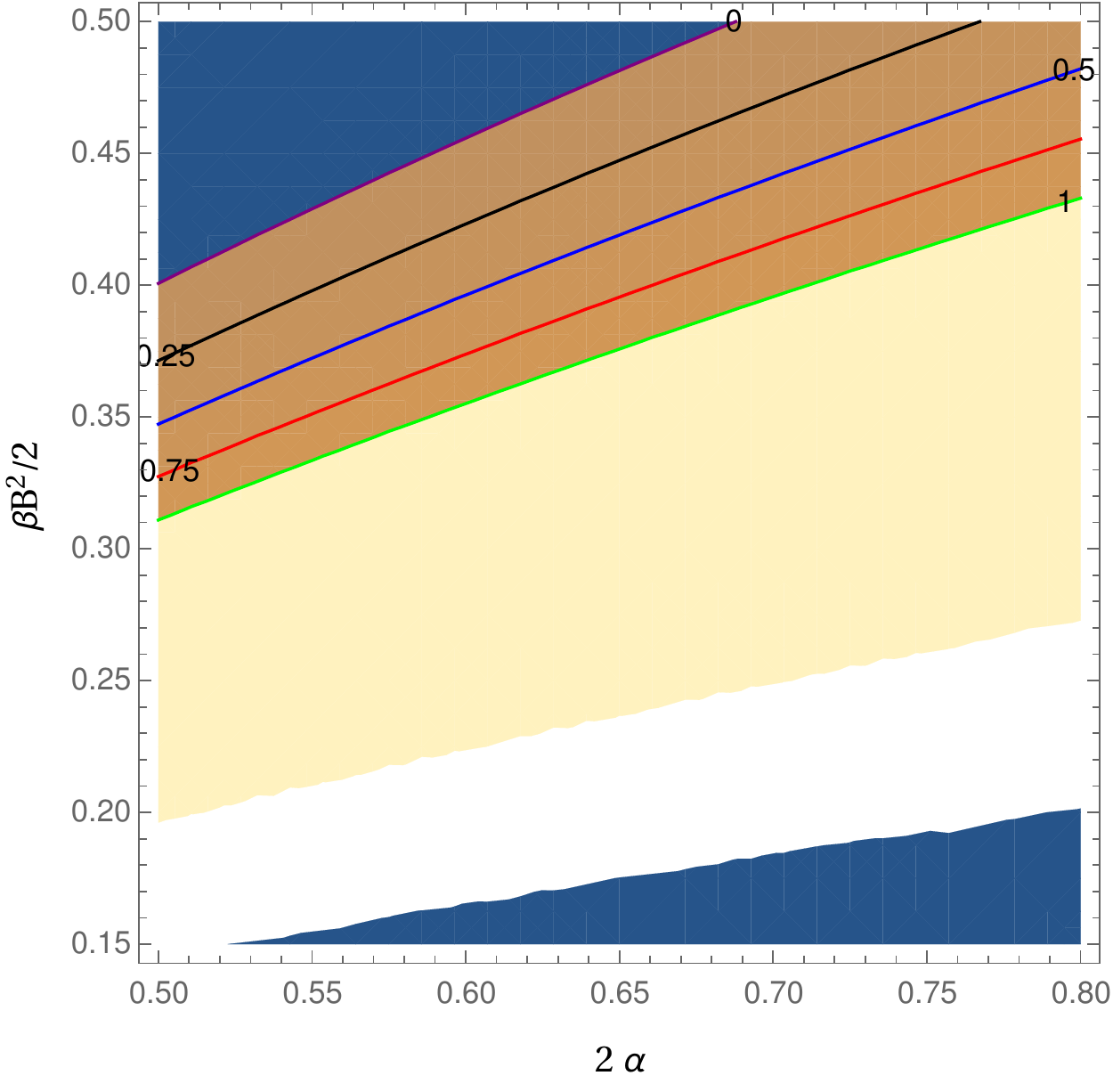}
% %\vspace*{-0.25in}
  \caption{{(Color online) On the left panel, the variation of $C_s^2 $ against $\beta B^2/2$ is shown, while on the right panel, the contour plots in plane of $\beta B^2/2$ and $2 \alpha$.}}
 \label{Plot1}
  \end{figure}

\begin{figure}[!h] 
 \centering
 \includegraphics[width=7.5cm]{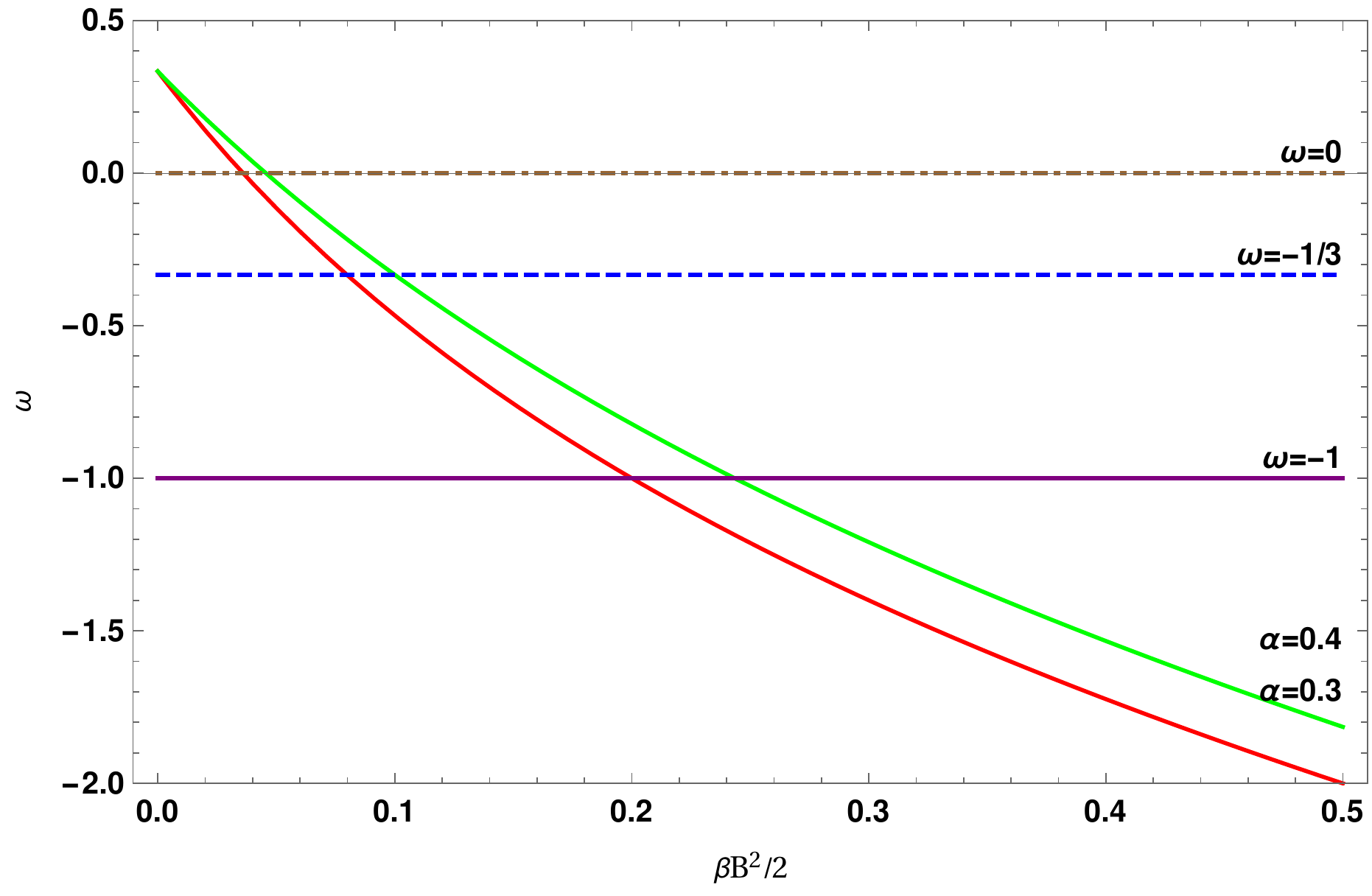}\hspace*{0.1in}
  \vspace*{-0.25in}
  \caption{{(Color online) Plot showing the variation of $\omega$ against $\beta B^2/2$.}}
 \label{Plot2}
  \end{figure}

\begin{figure}[!h]
\centering
 \includegraphics[width=8.5cm]{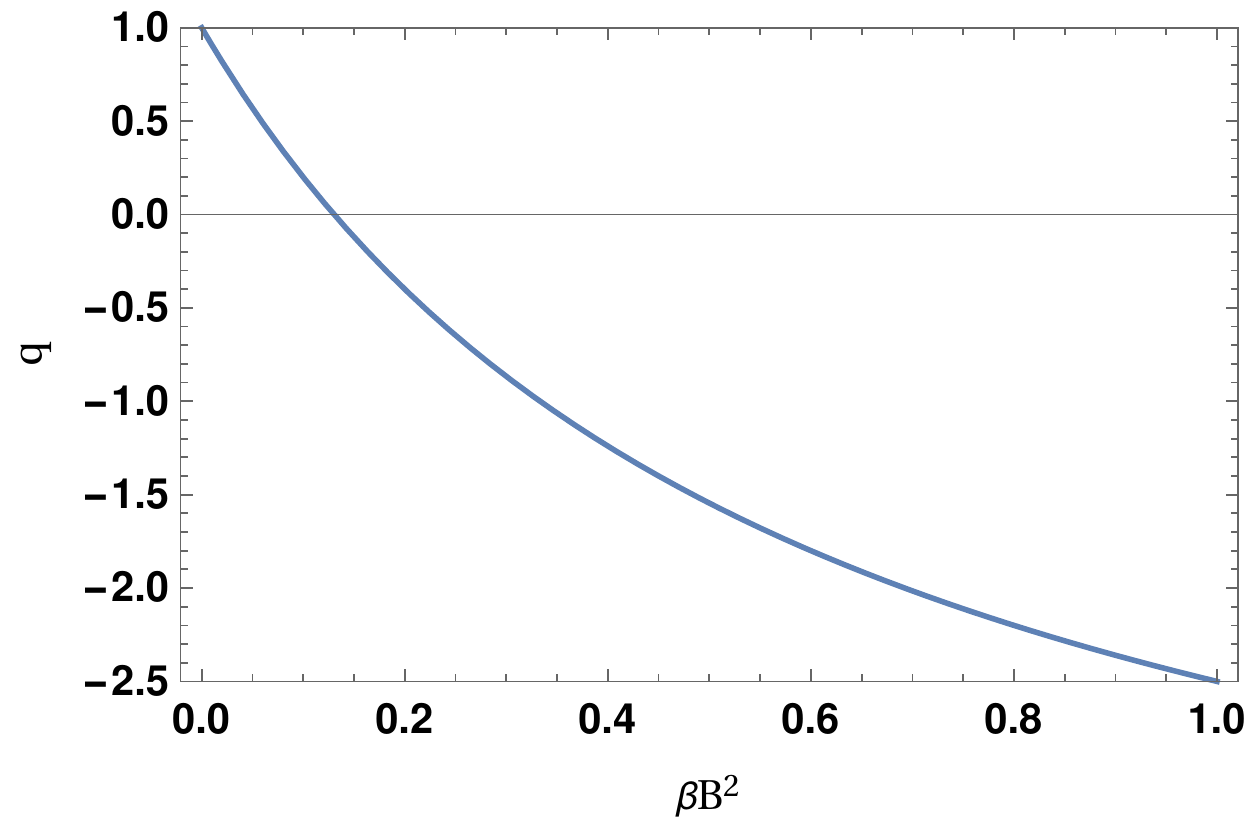}\hspace*{0.01in} 
\includegraphics[width=7.0cm]{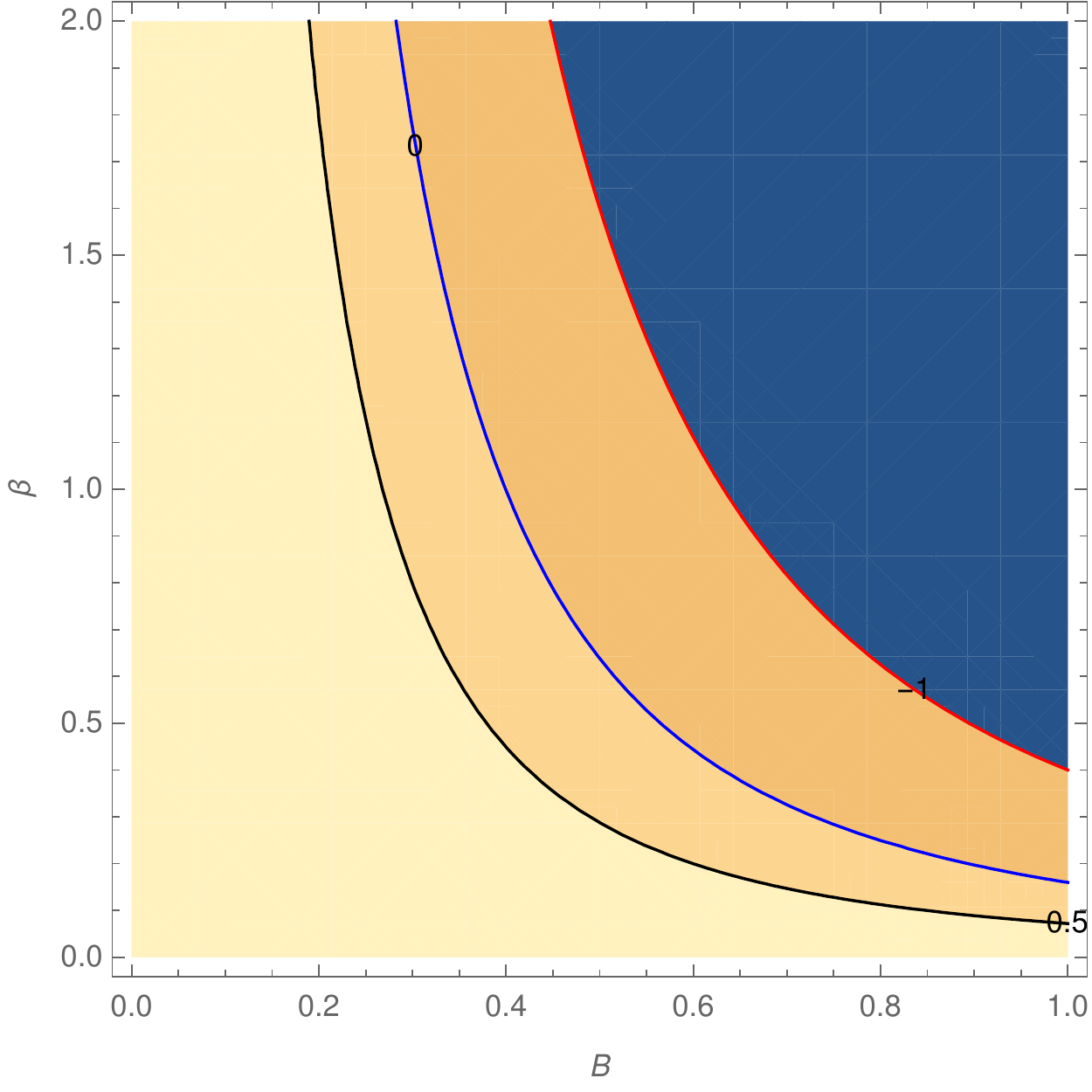}
 \caption{{(Color online) Plot showing the variation of $q$ against $\beta B^2$ corresponding to $\alpha = 0.3$.}}
\label{Plot3}
\end{figure}

We first investigate the causality and classical stability of the NLED model of inflationary expansion. The causality occurs if the speed of sound($C_s$) is less than the speed of light ($c$) i.e. $C_s (= dP_B/d\rho_B) \leq 1$ (here we set the speed of light $c = 1$ (natural unit)), while the classical stability requires $C_s^2 > 0$. From Eq.~(\ref{density}) and Eq.~(\ref{pressure}), we find  
\begin{equation}
 C_s^2=\frac{dP_B}{d\rho_B}=\frac{dP_B/d\mathcal{F}}{d\rho_B/d\mathcal{F}}=-1-\frac{2}{3}\beta B^2+\frac{8\alpha}{2\alpha+\beta B^2}+\frac{16\alpha-2\beta B^2(2\alpha+2)}{6\alpha\left(\frac{\beta B^2}{2}-1\right)+3\beta B^2\left(\frac{\beta B^2}{2}+1\right)}
\end{equation}
%%%%%%%%%%%%%
On the left of Fig.~(\ref{Plot1}), we have plotted $C_s^2$ against $\beta B^2/2$ for different $\alpha$ values, while  on the right the contour plots are shown in the plane of $\beta B^2/2$ and $2 \alpha$ corresponding to $C_s^2 = 0,~0.25,~0.5,~0.75$ and $1.0$.
%%%%%%%%%
%%%%%%%%%%%%5
On the left panel, for $0.25<\alpha<0.4$ we see that $0<C_s^2\leq1$. 
%So, the model is classically stable and causality is respected. In the rest of our analysis, we will consider $\alpha$ lying in this range.
%%%%%%%%%%%%%%5555
%  \begin{figure}[!h]
%  \centering
%   \includegraphics[width=6.5cm]{Figure2} 
% % %\vspace*{-0.25in}
%   \caption{{(Color online) Plot showing the variation of $\beta B^2/2 $ against $2\alpha$ for $C_s^2$ between 0 to 1 .}}
%  \label{Plot2}
%   \end{figure}
%%%%%%%%%%%%
% In Fig.~(\ref{Plot2}) we have shown some contour plots in the plane of $\beta B^2/2$ and $2 \alpha$ corresponding to $C_s^2 = 0,~0.25,~0.5,~0.75$ and $1.0$. 
While on the right panel, we see that $0 < C_s^2 < 1$  for  $0.25<\alpha<0.4$ and $0.6<\beta B^2<1$. This suggests that the  model is classically stable and it's causality is respected in some region of the parameter space. In the remaining part of our analysis, we will confine ourselves in these regions of $\alpha$ and $\beta B^2$. \\ 
%%%%%%%%%%%%%
In Fig.~(\ref{Plot2}), we have plotted the equation of state parameter $\omega$ against $\beta B^2/2$ for different $\alpha$
%%%%%%%%%%%%%
From the Fig.~(\ref{Plot2}), we find that Universe has large negative EoS for small a( large $\beta B^2$) and it crosses $\omega=-1$ at $\beta B^2=0.4(0.5)$ corresponding to $\alpha=0.3(0.4)$, while for $\omega=-1/3$ and $\alpha=0.3(0.4)$, we find $\beta B^2=0.15 (0.2)$, respectively. Universe changes from acceleration to deceleration phase.In the limit $\beta B^2\rightarrow 0$, $\omega$ approaches 1/3( radiation dominated Universe). Next, we define the deceleration parameter($q$) as 
\begin{equation}
 q= - \frac{a \ddot{a}}{\dot{a}^2} = \frac{1}{2}(1+3\omega)=\frac{2\alpha-3\beta B^2-2\alpha\beta B^2-\beta^2B^4}{2\alpha+\beta B^2}
\end{equation}
%%%%%%%%%%%%%%%%
In Fig.~(\ref{Plot3}), on the left panel, we have shown how the deceleration parameter $q$ varies with $\beta B^2$ for $\alpha = 0.3$, while on the right panel, we have made a contour plot in the $\beta~-~B$ plane corresponding to $q=0.5, 0$ and $-1$ (which corresponds to $\omega = 0,~-1/3$ and $-1$), respectively.   
%%%%%%%%%%%%%%%%%%%%
 %%%%%%%%%%%%%%%%%%%%%
From the Fig.~(\ref{Plot3}), we find that the deceleration parameter $q$ remains negative so long $\beta B^2 \geq 0.28$ (for $\alpha = 0.3$) which implies the accelerating phase of the universe expansion.

\begin{figure}[!h]
\centering
 \includegraphics[width=7.0cm]{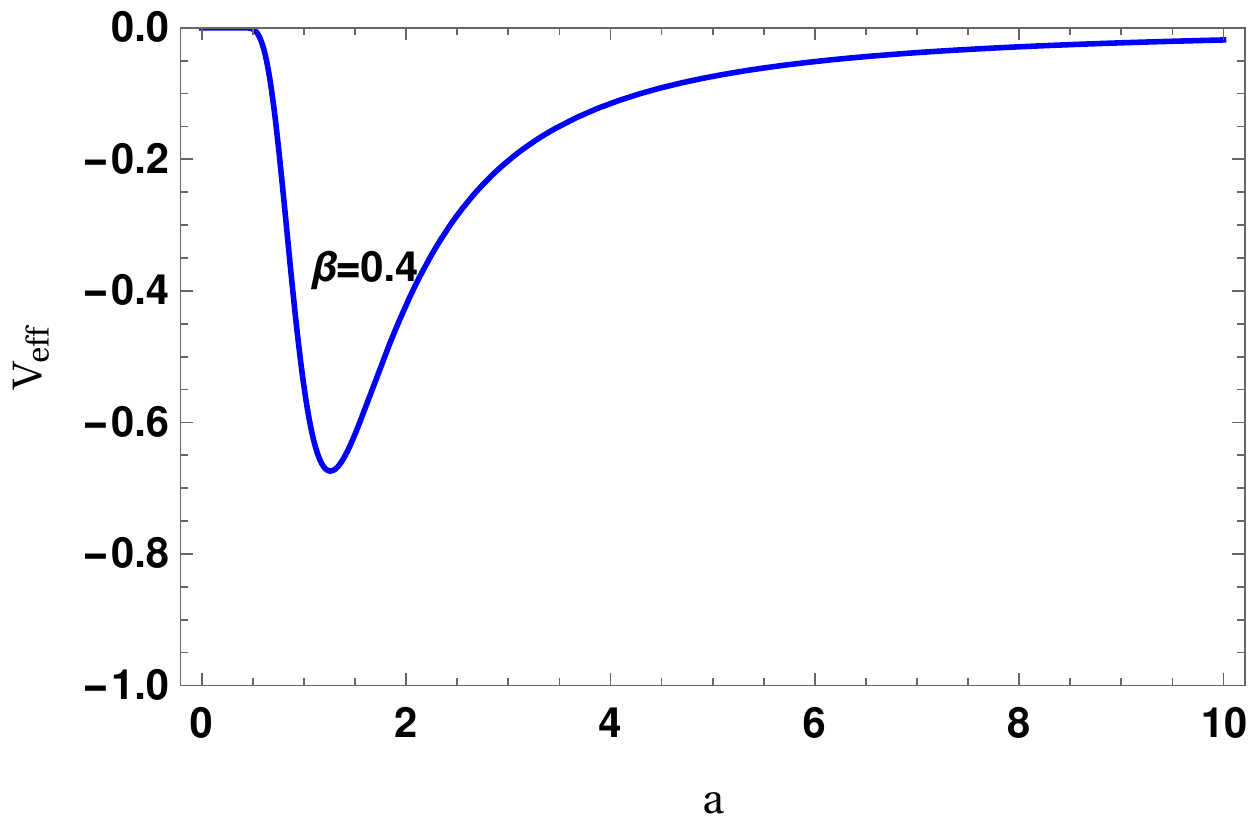}
\hspace*{0.01in} \includegraphics[width=6cm]{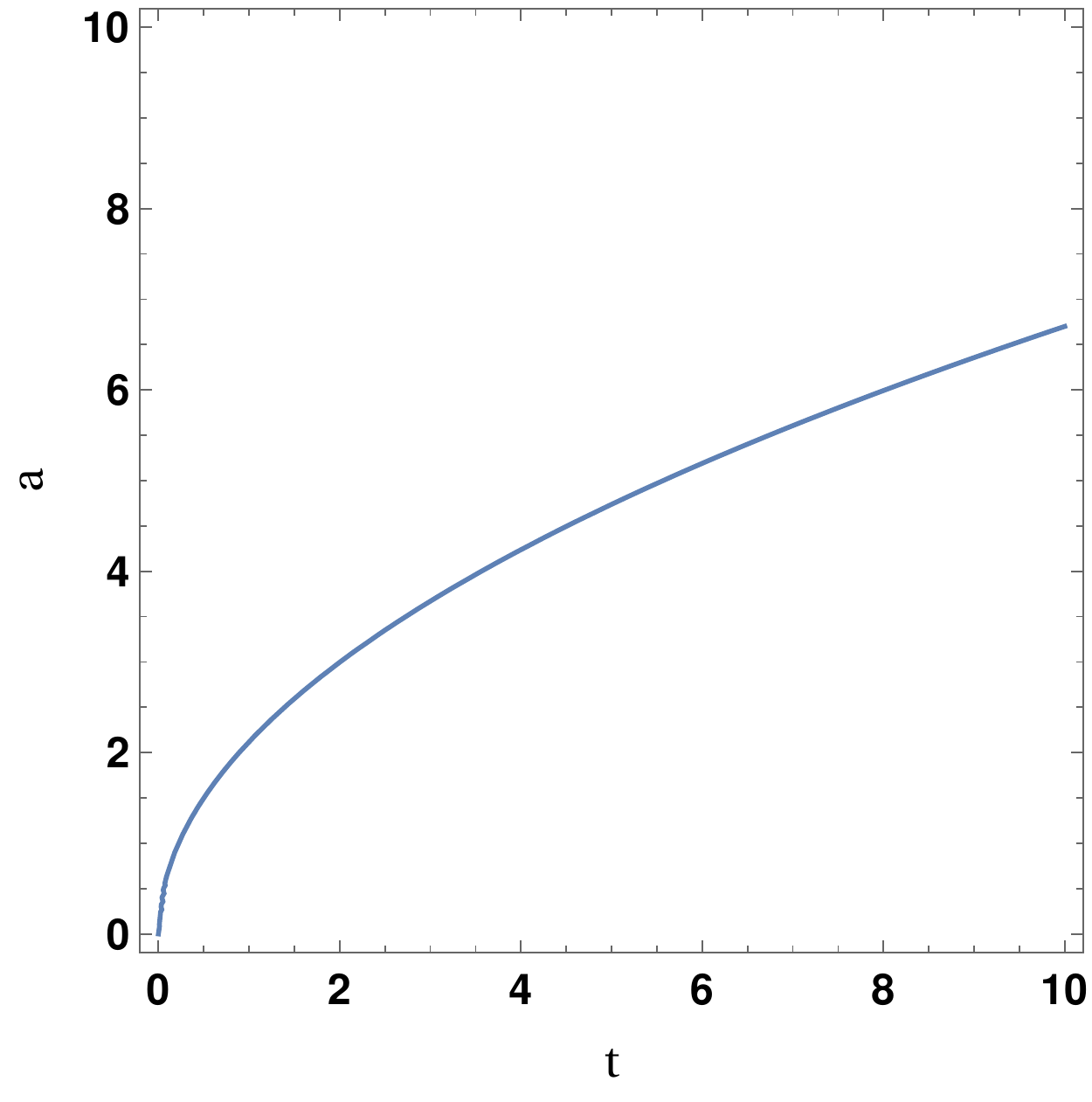}
 \caption{{(Color online) On the left panel, plot showing the variation of $V_{eff}$ against $a$ for $\alpha=0.3$ and $\beta=0.4$, while on the right panel, $a(t)$ is plotted against $t$.}}
\label{Plot4}
 \end{figure}

\noindent From the velocity (Friedmann) equation (for $k=0$ case(flat universe)), we find  
\begin{equation}
 H^2 = \left(\frac{\dot a}{a}\right)^2=\frac{\rho_B}{3} \label{velocity}
\end{equation}
we obtain the equation which shows the conservation of energy with effective potential $V_{eff}(a)$,
\begin{equation}
 \dot{a^2}+V_{eff}(a)=0 \label{veff}
\end{equation}
In above, $V_{eff}(a)=-\frac{B_0^2}{6a^2}e^{-\beta B_0^2/2a^4}\left(\frac{\beta B_0^2}{2a^4}+\alpha\right)^{-2}$ where we have used $B = B_0/a^2$. In Fig.~(\ref{Plot4}), on the left panel, we have plotted $V_{eff}(a)$ as a function of $a$ for $\beta = 0.4$ and $\alpha = 0.3$. 
% Note that for $\alpha = 0.3$, $\beta B^2 \to 0.4$ as $\omega = -1$ (also follows from $C_s^2$ plot (for $\alpha = 0.3$) corresponding to $\omega = -1$). 
% We have chosen this value of $\beta$ while plotting $V_{eff}$ against $a$. 
%%%%%%%%%%%%% 
%%%%%%%%%%
We see that for the entire range of values of $a$, $V_{eff}$ always remain negative, which suggests that 
$\dot{a}^2 > 0$ throughout.
For small $B=\frac{B_0}{a^2}$ due to large $a$, we obtain (from Eq.~\ref{veff}) $\dot{a^2}=\frac{B_0^2}{6\alpha^2 a^2}$. Integrating we get, $t-t_0=\sqrt{\frac{3 \alpha^2}{2 B_0^2}} ~a^2$ where $t_0$ is the integration constant. Setting $t_0=0$, the present time and $B_0 = 1$ (the present day normalized magnetic field value), we find (with $\alpha = 1$) 
\begin{equation}
 a = \left(\frac{2}{3}\right)^{1/4} \left(\frac{t}{\alpha}\right)^{1/2}
\end{equation}
A plot showing the variation of $a(t)$ against $t$ is shown on the right panel of Fig. \ref{Plot4} and the behaviour of $a(t)$ as a function of $t$ is as per the normal electrodynamics, since in the limit $\alpha \to 1$, ${\mathcal{L}}_{NED} \to - {\mathcal{F}}$, the usual QED lagrangian. Note that at $t=0, a_0=0$, the function $a_0$ is the radius of Universe which shows that Universe begins from the zero point.
Similarly, the acceleration (the Raychaudhury) equation gives,  
\begin{equation}
 \ddot{a}+ W_{eff}=0 \label{acceleration}
\end{equation}
Where $W_{eff}=-\frac{B_0^2}{12a^3}e^{-\beta B_)^2/2a^4}\left(\alpha+\frac{\beta B_0^2}{2a^4}\right)^{-3}\left[\frac{\beta^2B_)^4}{a^8}+\frac{\beta B_0^2}{a^4}\left(2\alpha+3\right)-2\alpha\right]$.
%%%
%%%%%%%%%%%%%%%%%%%%
%%%%%%%%%%%%%%5
  \begin{figure}[!h]
  \centering
   \includegraphics[width=7.0cm]{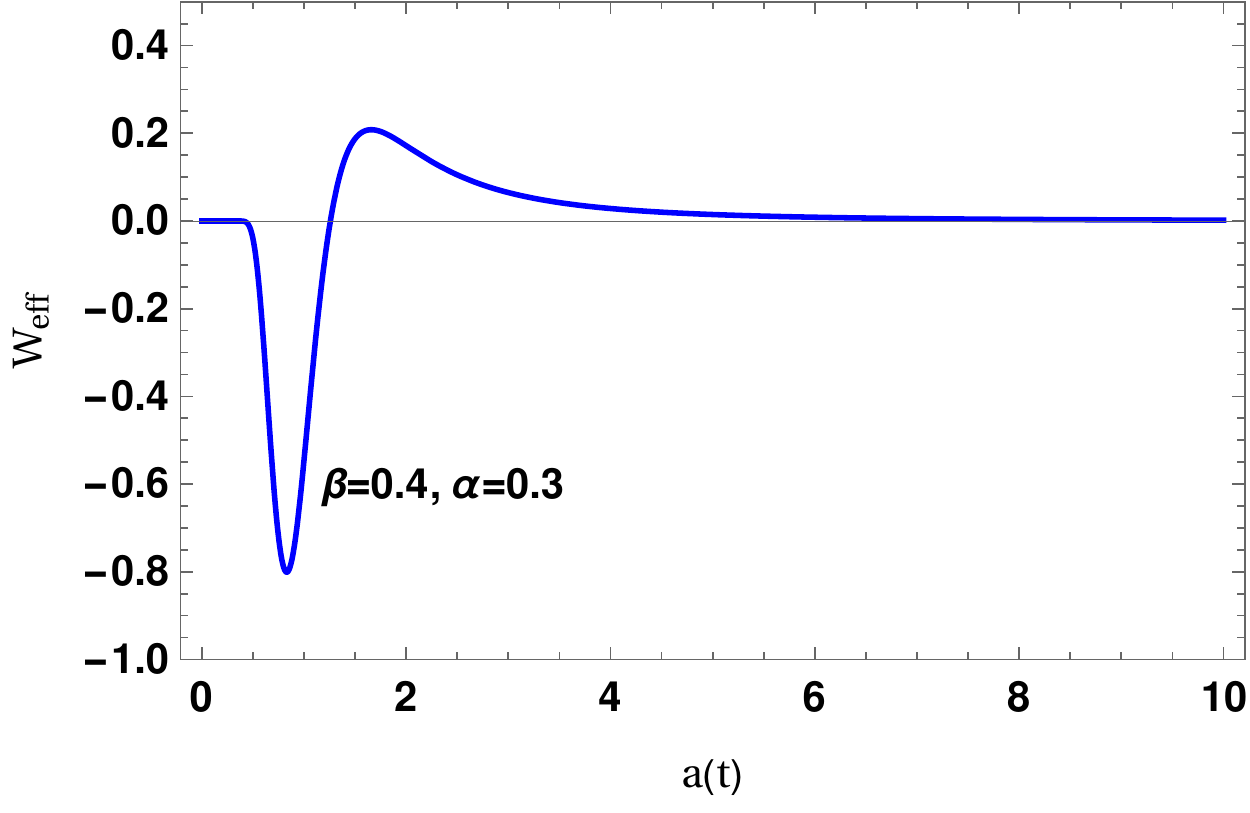}
 \hspace*{0.01in} \includegraphics[width=7.0cm]{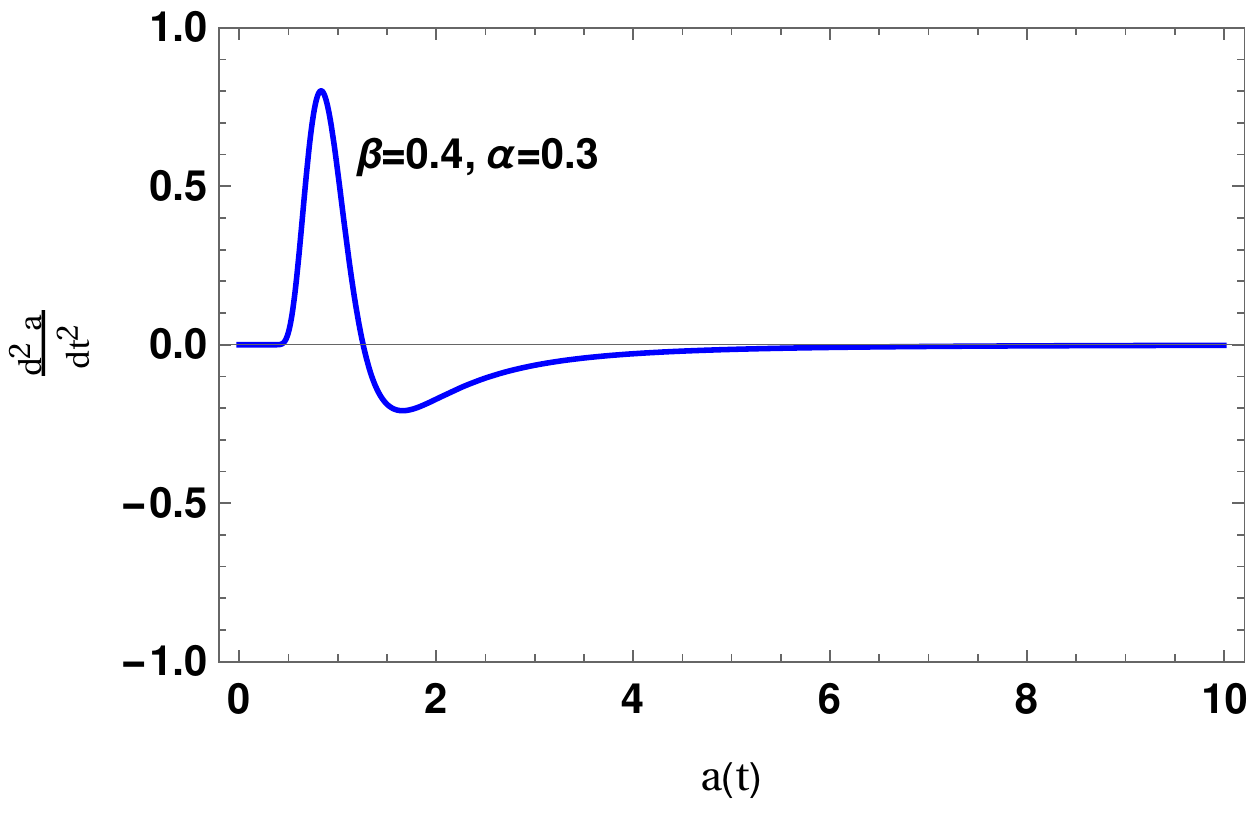}
   \caption{{(Color online) Plots showing the variation of $V_{eff}$ against $a$(on the left panel) and $\ddot{a}$ against $a(t)$(on the right panel) corresponding to $\alpha=0.3, \beta=0.4$.}}
  \label{Plot5}
  \end{figure}
%%%%%%%%%%%%%%%%
% \begin{figure}
% \centering
%  \includegraphics[width=6.5cm]{Figure7}
% \hspace*{0.01in} %\includegraphics[width=6.5cm]{Figure1b}
%  \caption{{(Color online) Plot showing the variation of $\ddot{a}$ against $a$.}}
% \label{Plot7}
%  \end{figure}
 %%%%%%%%%%%%%%%5
On the left panel of Fig.~(\ref{Plot5}), we have plotted $W_{eff}$ against  $a$, while on the right panel $\ddot{a}(t)$ is shown as a function of $a(t)$ corresponding to $\alpha=0.3$ and $\beta=0.4$. We see that as long as $W_{eff}$ remains negative, the universe accelerates i.e. $\ddot{a}(t) > 0$.
%%%%%%%%%%%%%%%%
%%%%%%%%%%%%%
Accordingly, one finds the magnetic field which evolves as
 \begin{equation}
  B(t)= \frac{B_0}{a^2} = \frac{\alpha \sqrt{6}}{2}\frac{1}{t} \label{BB}
 \end{equation}
%%%%%%%%%%%%%%
 \begin{figure}[htb]
\centering
 \includegraphics[width=6.5cm]{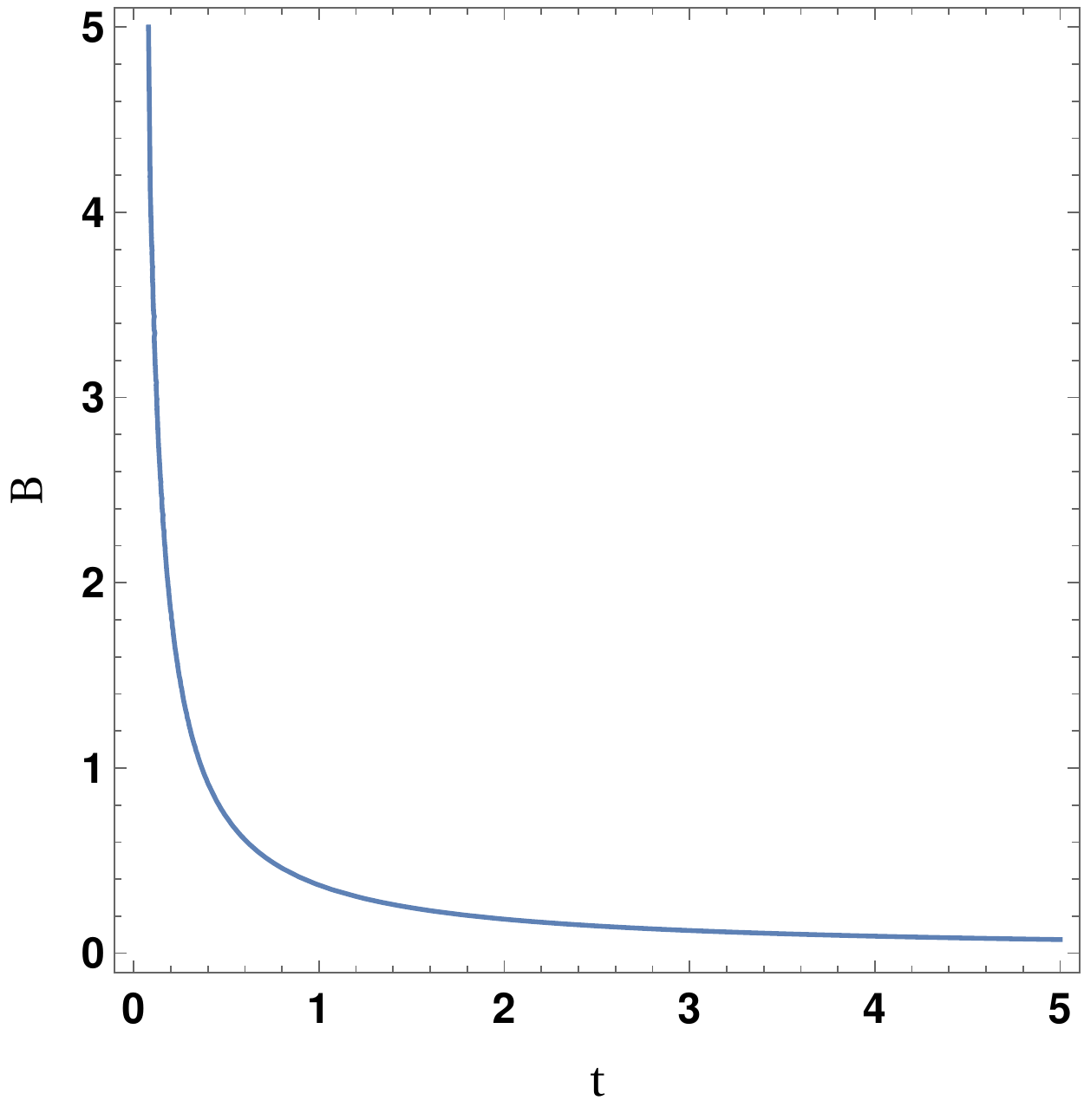}
\hspace*{0.01in} \includegraphics[width=6.5cm]{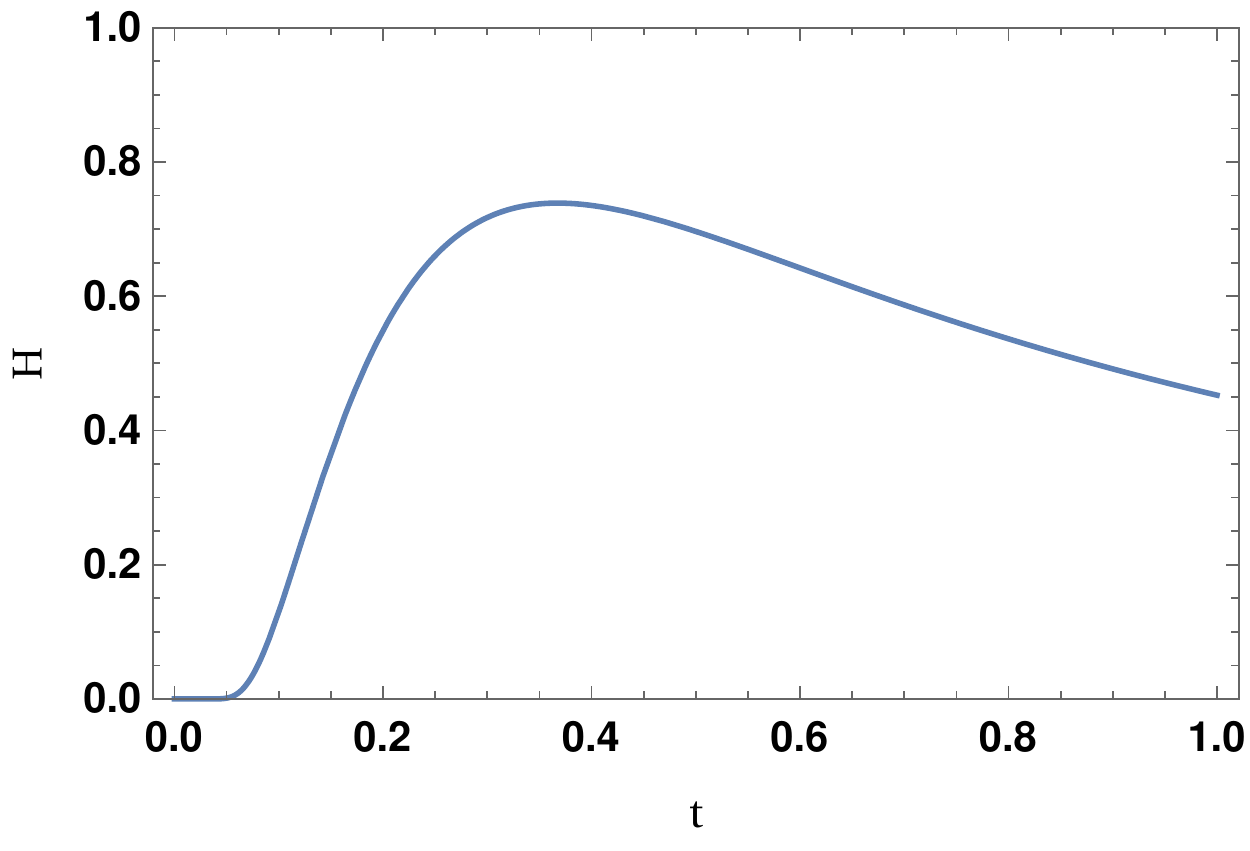}
 \caption{{(Color online) On the left, $B(t)$ is plotted against $t$, while on the right the Hubble parameter $H(t)$ is plotted as a function of $t$ with $\alpha=0.3$.}}
\label{Plot6}
 \end{figure}
 %%%%%%%%%%%%%%%%%%%55
 Using Eq.~(\ref{velocity}), one finds Hubble parameter $H$ varies with $t$ as
 \begin{equation}
  H=\frac{\alpha e^{-\left(\frac{3\alpha^2\beta}{8t^2}\right)}}{t\left(2\alpha+\frac{3\alpha^2 \beta}{2t^2}\right)} \to \frac{1}{2 t} 
 \end{equation}
% %%%%%%%%%%%%
in the limit $\alpha \to 1$ and $\beta \to 0$. The plots showing the variation of $B(t)$ against $t$ and $H(t)$ against $t$ are shown in 
Fig.\ref{Plot6}.  \\

\begin{figure}[htb]
    \centering
    {\includegraphics[width=7cm]{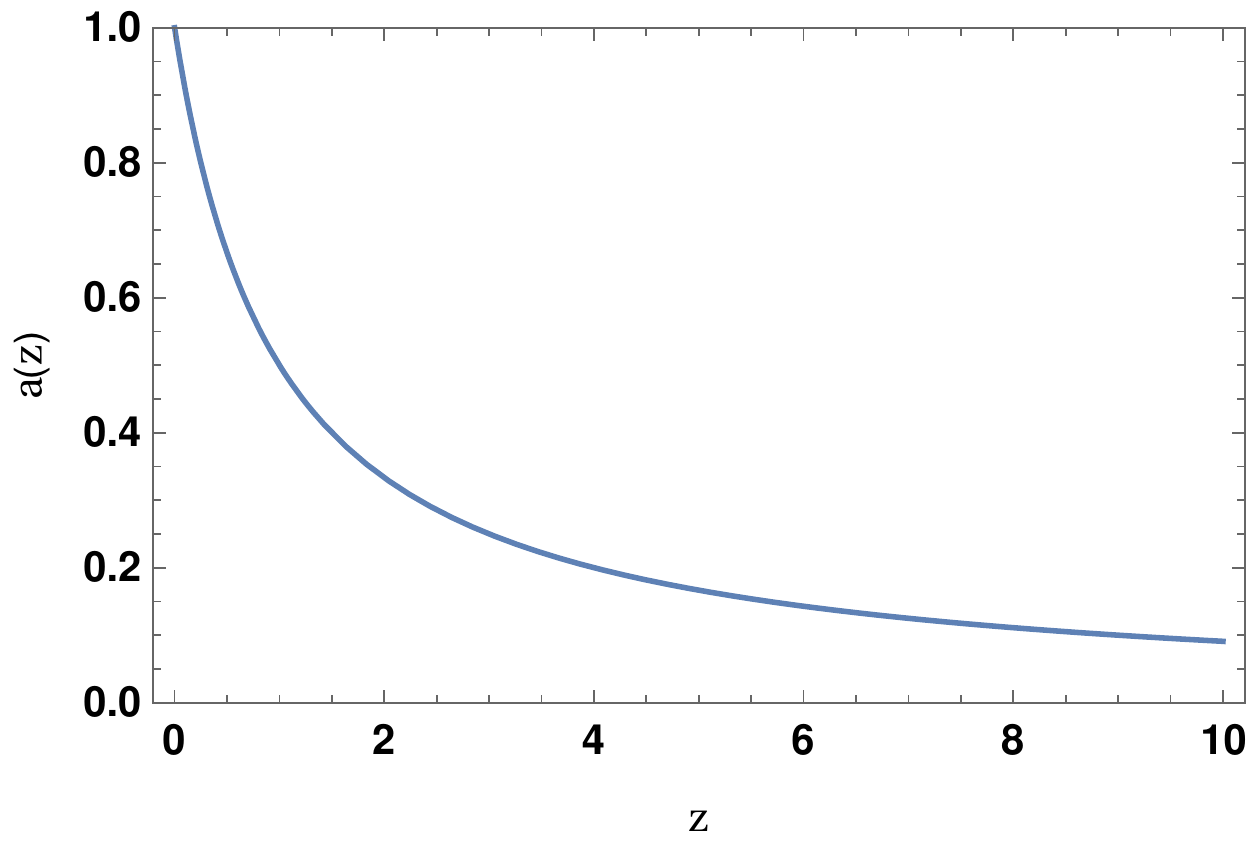}}
%     \qquad
%   {\includegraphics[width=7cm]{Figure8b}}
    \caption{{(Color online)Plot showing the variation of $a(z)$ against $z$.}}   \label{Plot8}
\end{figure}

\begin{figure}[htb]
    \centering
    {\includegraphics[width=5cm]{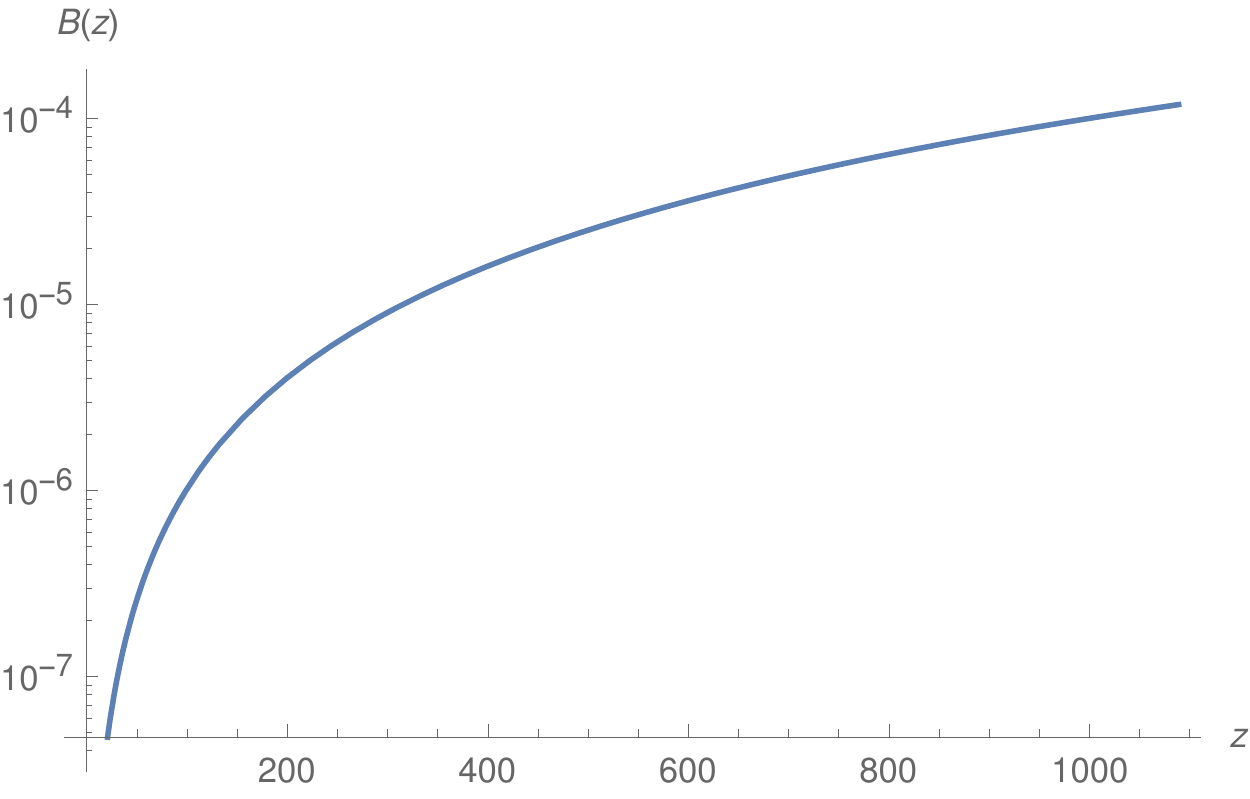}}
    \qquad
  {\includegraphics[width=5cm]{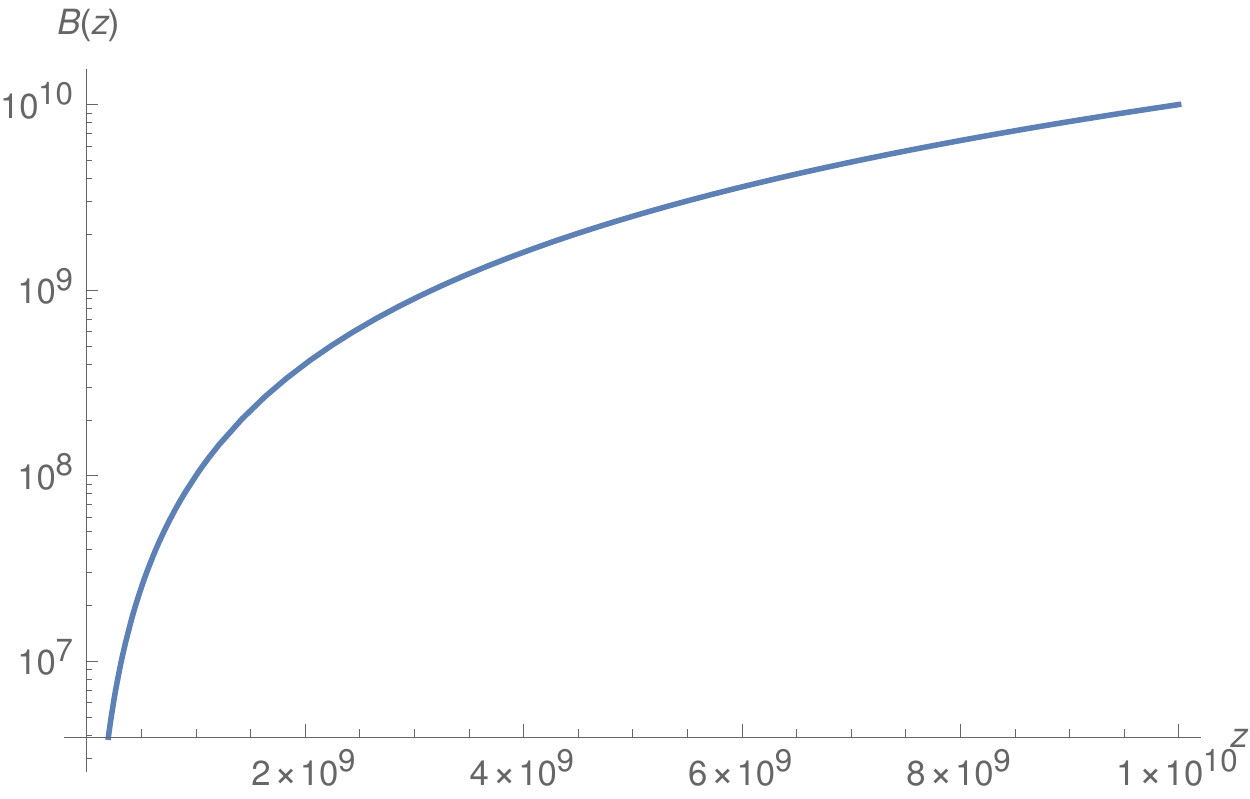}}
    \qquad
  {\includegraphics[width=5cm]{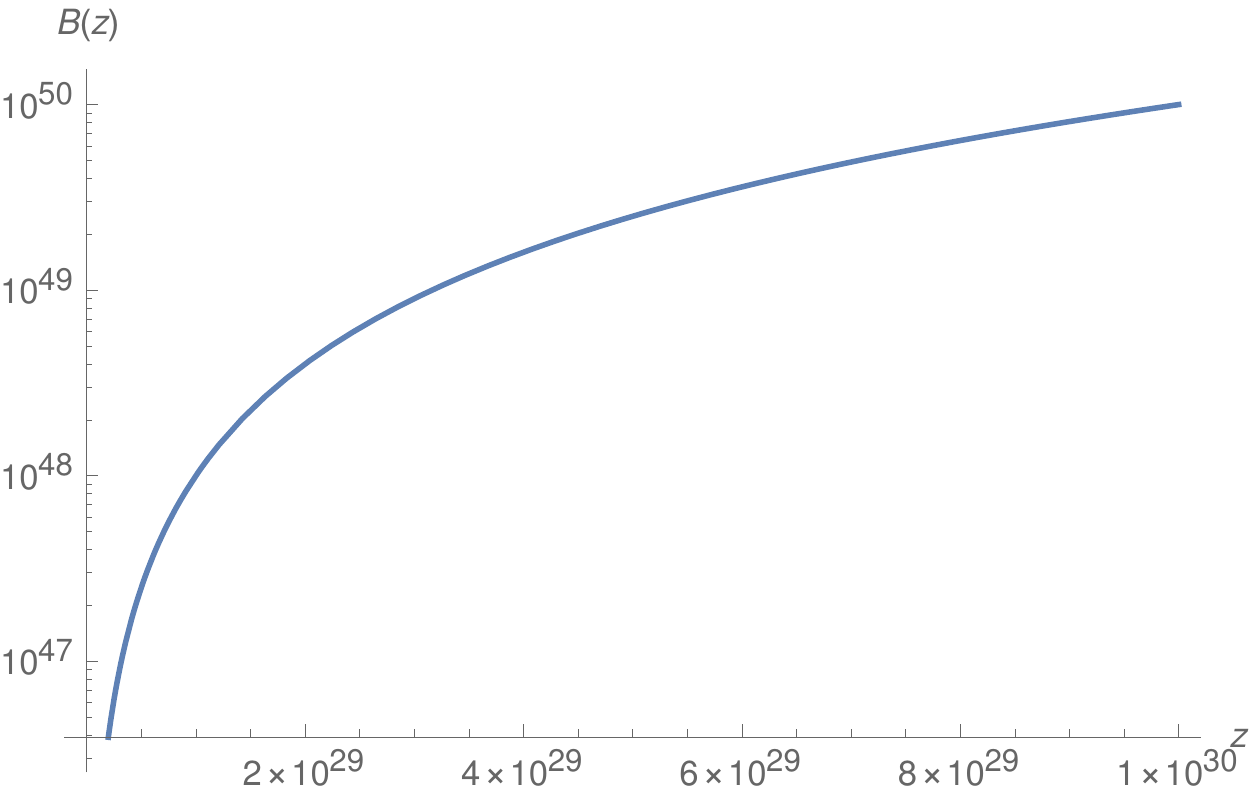}}
    \caption{{(Color online)Plots showing the variation of $B(z)$ (in Gauss(G)) against $z$. On the extreme left, $z$ varies from $0$ to $1089$, in the case of middle one $z$ varies from $1089$ to $10^{10}$, whereas on the extreme right, $z$ varies from $10^{10}$ to $10^{30}$.} }   \label{Plot9}
\end{figure}

Also during inflation, one finds $\dot{\rho_B} = 0$ (from the equation of continuity) i.e. $\rho_B$ = constant (as $\rho_B + P_B = 0$) and this gives   
\begin{equation}
 \beta^2B^4+\beta B^2(2+2\alpha)-4\alpha=0 \label{EOC}
\end{equation}
%%
%%%%%%%%%%%%%%
\begin{figure}[htb]
    \centering
    {\includegraphics[width=7cm]{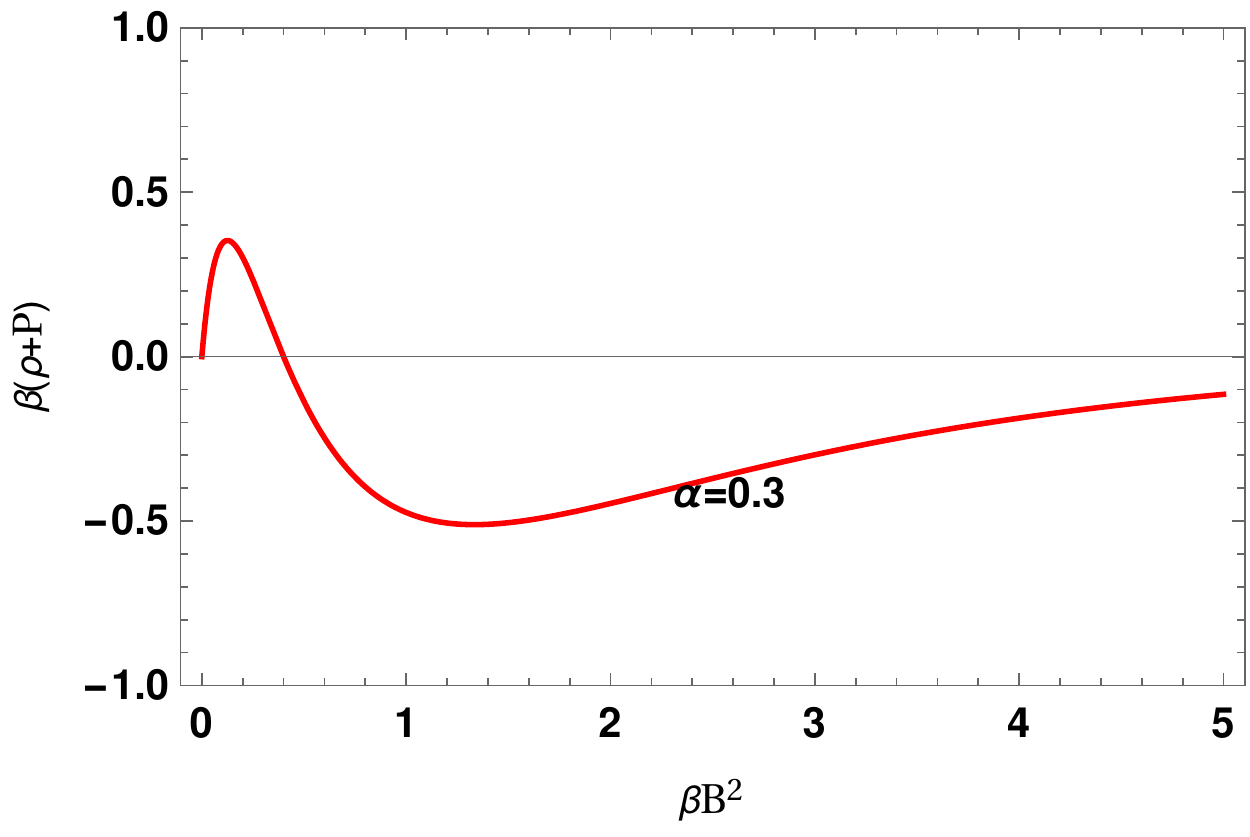}}%
    \qquad
  {\includegraphics[width=7cm]{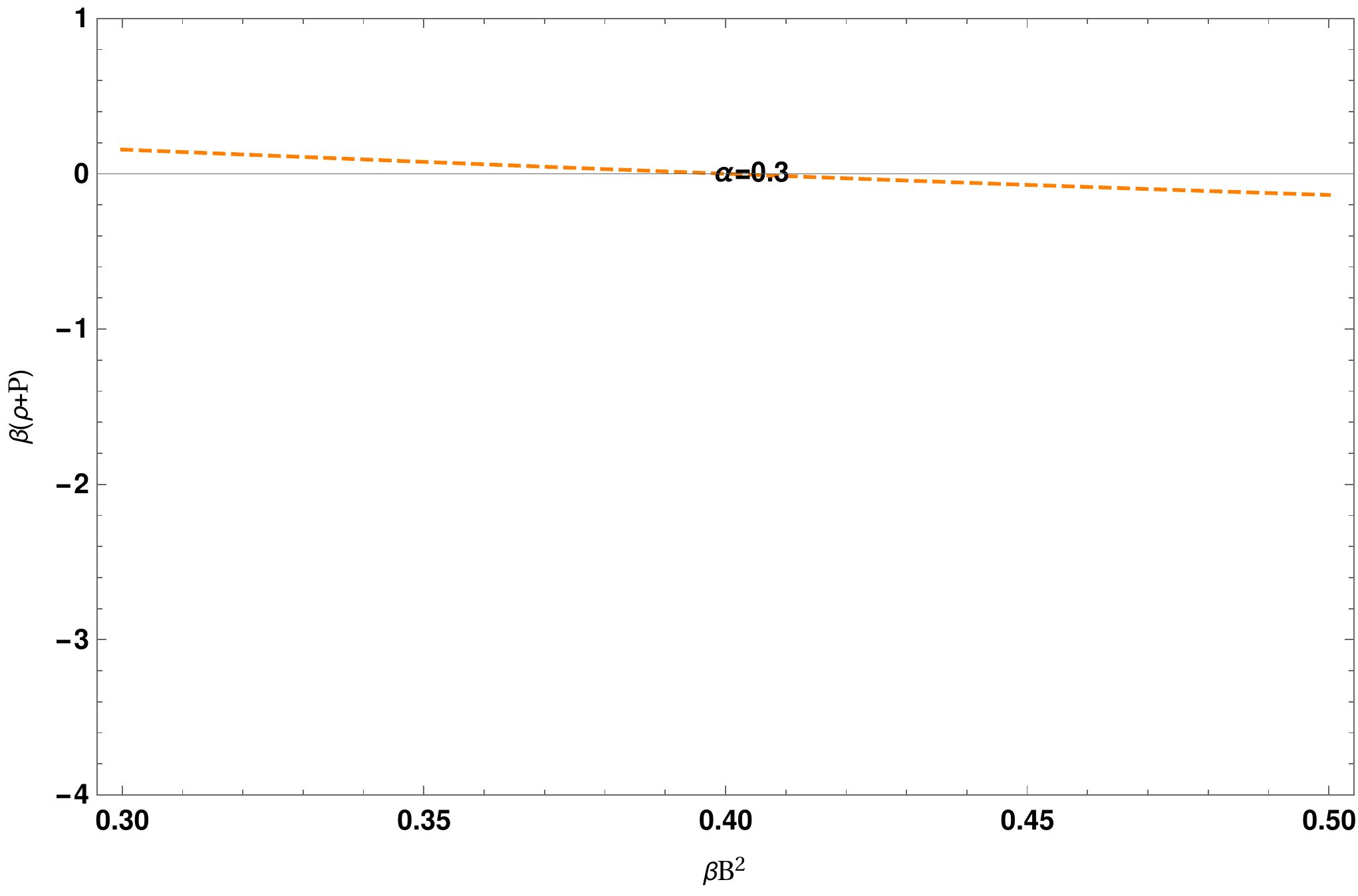}}%
    \caption{{(Color online) Plot showing the variation of $\beta(\rho+3P)$ against $\beta B^2$ with $\alpha=0.3$.}}   \label{Plot7}%
\end{figure}
%%%%%%%
\noindent From the above equation, we get $\beta B^2=0.4$ for $\alpha=0.3$, which is found to be consistent with Fig.~(\ref{Plot7}) where we have plotted $\beta(\rho_B + P_B)$ against $\beta B^2$. Using this $\beta B^2=0.4$ value in Eq.~(\ref{EOC}), one finds the maximum energy density $\rho_B$ occurs at 
\[\rho_B = \frac{2 \beta B^2 e^{-\beta B^2/2}}{\beta (2 \alpha + \beta B^2)^2} = 0.65/\beta = \rho^{max}_B ~\to \beta \rho^{max}_B = 0.65 \] where we have set $\alpha = 0.3$. Note that $\rho_B (= \rho^{max}_B)$ remains constant during inflation. This energy density $\rho_{max}$  gives the value of the magnetic field $B \simeq \sqrt{\frac{0.4}{\beta}} = \sqrt{\frac{0.4\rho^{max}_B}{0.65}}$ for the inflationary phase. 
In a typical inflationary model (e.g. chaotic, hybrid, natural) where the reheating temperature is $T_{reh} \sim \rho_{inf}^{1/4}  \sim 10^{16}~\rm{GeV}$, i.e. at grand unification scale, the energy density during inflation is taken to be $\rho_{inf}(=\rho^{max}_B) = 10^{64}~\rm{GeV^4}$.  Accordingly, we find the magnetic field to be at the beginning,
\begin{equation}
 B_{start}=\sqrt{\frac{0.4 (=\rho^{max}_B)}{0.65}}=0.79\times10^{32}\frac{1}{2\times10^{-20}}G=0.395\times 10^{52} \sim 4 \times 10^{51}G
\end{equation}
% Now the e-fold number($N$) required to produce inflation is defined as 
%  \begin{equation}
%       N=\ln\frac{a_{end}}{a_{start}} =\ln{\sqrt{\frac{B_{start}}{{B_{end}}}}}
%   \end{equation}
% Considering, $B_0 \simeq B_{end} = 10^{-10}G $, the e-fold number $N=\frac{1}{2}\times \ln({4 \times 10^{61}}) = 71$ which agrees with the experimental prediction.
%%%%%%%%%%%%%%%
In Fig.~(\ref{Plot8}), we have plotted the scale factor $a$ (where $a=\frac{1}{1+z}$) as a function of the redshift factor $z$. We see that as $z\rightarrow 0$, the scale factor $a$ increases to a very large value corresponding to the current universe, whereas as 
$z \to$ large value (corresponding to early day universe), $a\rightarrow 0$. 
%%%%%%%%%%
%%%%%%%%%%%%%%
% In Fig.~(\ref{Plot8}), we have plotted the scale factor $a$ (where $a=\frac{1}{1+z}$) as a function of the redshift factor $z$. We see that as $z\rightarrow 0$, the scale factor $a$ increases to a very large value corresponding to the current universe), whereas, corresponding to early day universe, as 
% $z \to$ large value, $a\rightarrow 0$. 
In Fig.~(\ref{Plot9}), we have shown how the magnetic field $B$ evolves with $z$. We see that as $z$ increases from $0$(present day value) to $1089$ (the redshift value corresponding to the event of first CMBR formation) and further to extremely high value $10^{29}$ (the redshift value corresponding to the event inflation), $B$ varies from a very small value $10^{-10}$ G (present day value) to $10^{-4}$ G (the value at the time of CMBR formation) and then to $\sim 10^{51}$ G, the very $B$ value necessary to trigger inflation. 
%%%%%%%%%%%%%%%
%%%%%%%%%%%%%%
% \\ Note that $B(z)$ evolves with the redshift factor($z$) as $B = B_0(1+z)^2$. Noting that $(1+z)^2 = \frac{1}{a^2} = \frac{B(t)}{B_0}= \frac{1}{B_0} \frac{\alpha \sqrt{6}}{2 t}$, one finds as $t \to t_{in}$(where $t_{in} = 10^{-36}$ sec, the time when inflation starts), $z\approx 10^{23}$ at the time when inflation starts and the magnetic field necessary to trigger the inflation is found to be $B = 10^{-10}(1+10^{23})^2=10^{50}G$ (with $B_0 = 10^{-10}$ Gauss, the present day magnetic field). Note that this agrees with our earlier estimate. 
% We see that as $z$ increases from $0$ to extremely high value $10^{29}$, $B$ varies from a very small value $10^{-10}$ G (present day value) to a very large value $\sim 10^{51}$ G, which is the required $B$ value to drive inflation.  
%  Looking at the evolution of the universe temperature, one can also make an estimate of the magnetic field necessary to drive the inflation. 
 Noting, if inflation scale is determined by the reheating temperature $T = 10^{16}~\rm{GeV} = 5 \times 10^{29} \rm{cm^{-1}} \approx 1/a$, one finds $a = 0.2 \times 10^{-29} \rm{cm}$, and $z = 5 \times 10^{29}$ which corresponds to  $B(t = t_{in} = 10^{-36}~\rm{sec}) = B_0 (1 + z)^2 \approx 3 \times 10^{49}$ G, a value close to our earlier estimate of $B \sim 10^{51}$ G.\\
%%%%
{\bf e-fold number(N)}:~ The e-fold number($N$) in terms of the magnetic field $B$ is defined by     
 \begin{equation}
      N=\ln\frac{a_{end}}{a_{start}} =\ln{\sqrt{\frac{B_{start}}{{B_{end}}}}}
  \end{equation}
Considering, $B_0 \simeq B_{end} = 10^{-10}$ G (corresponding to $z=0$) and $B_{start} = 4 \times 10^{51}$ G,  we find $N=\frac{1}{2}\times \ln({4 \times 10^{61}}) = 71$ (whereas for $B_{start} = 3 \times 10^{49}$ G, we find $N = 69$). Also, if we consider $B_{end} = 10^{-4}$ G (corresponding to $z = 1089$) and $B_{start} = 4 \times 10^{51}$ G,  we find $N=\frac{1}{2}\times \ln({4 \times 10^{55}}) = 64$ (whereas for $B_{start} = 3 \times 10^{49}$ Gauss, we find $N = 62$). This agrees quite well with the experimental result.

\section{Conclusion:}
In this paper we have considered a new kind of NLED field which acts as a source of gravity and can accelerate the universe to accelerate during the inflationary era.  We have studied the isotropic and homogeneous magnetic universe where the newly proposed NLED lagrangian is charecterized by two paremetrs $\alpha$ (dimensionaless parameter) and $\beta$ (dimensionful parameter). We have investigated the classical stability and causality aspects of our model of inflationary expansion by demanding that the speed of the sound wave  $C_s^2 > 0$ and $C_s \le 1$. We found that $0 < C_s^2 < 1$ for $0.25 \le \alpha \le 0.4$ and $0.6 \le \beta B^2 \le 1$. A plot of $\omega$(equation of state parameter $=P/\rho$) against $\beta B^2/2$ for $\alpha = 0.3$ suggests that $0.2 \le \beta B^2 \le 0.4$ provided $-1/3 \le \omega \le -1$. The deceleration parameter($q$) study also suggests that $q < 0$ (i.e. $\omega < -1/3$ with  $q = \frac{1}{2}(1 + 3 \omega)$) and hence $\ddot{a} > 0$ ( the universe is accelerating) provided $\beta B^2 \ge 0.13$.
During inflation, the energy density $\rho_B$ is found to be maximum and is given by $\rho_B^{max} = 0.65/\beta$. The magnetic field necessary to trigger the inflation, is found to be $B \simeq \sqrt{\frac{0.4 \rho_B^{max}}{0.65}} = 4 \times 10^{51}~{\rm Gauss}$, where $\rho_{B}^{max}(\sim \rho_{inf}) =10^{64}~{\rm GeV}^4 $ is the energy density of the universe during inflation and we found that the field $B$ decreases with time $t$. Our model also predicts the e-fold number $N = 71(64)$ that the magnetic field at the end of inflation is about $B = 10^{-10}~(10^{-4})$ Gauss corresponding to $z=0(1000)$ and this agrees quite well with the experimental prediction of the e-fold number. We found that our model describes several aspects of inflationary cosmology and is classically stable in a range of parameters($\alpha, \beta$) space. 

% , assuming that the reheating occurs at the GUT scale i.e. $T_{reh} \sim \rho^{1/4}_{inf} \sim 10^{16}~\rm{GeV}$, we find the magnetic field at the beginning of inflation about $10^{51}~\rm{G}$. We also predict the e-fold number $N( \sim ln(\frac{\sqrt{B_0}}{H_0}) + ln(\frac{H_i}{\sqrt{B_f}})) = 69$ with $B_0 \sim 10^{-10}$ G (magnetic field at the end of inflation). From $C_s^2$ vs $\beta B^2/2$ plot Fig.~(\ref{Plot1}) we have found out that classical stability holds for $0.25<\alpha<0.4$ and $0.3<\beta B^2<1$. So this lagrangian model has good range of classical stable region.

\section{Acknowledgement}
PS thanks to Department of Science and Technology, Government of India for the Inspire fellowship (No. DST/INSPIRE Fellowship/2017/IF170807). The work of PKD and GCS is supported by CSIR Grant No.25(0260)/17/EMR-II
%%%%%%%%%%%%%%%%%%%%%5
%%%%%%%%%%%%%%%

%%%%%%%%%%%%%%%%%%%
%%
%e-Print: arXiv:1705.01455v1[gr-qc].
% %
% \bibitem{Dimopoulos}K.~ Dimopoulos, {\it Astrophys.Space Sci.Libr.} {\bf 276} (2002) 53 - 60.
% %
% \bibitem{heeck}J.~Heeck,{\it Introduction to Inflation}, \url{https://www.mpi-hd.mpg.de/lin/events/group_seminar/inflation/heeck.pdf}
% 
% \bibitem{sirichai}Sirichai Chongchitnan (Hull U.),{\it Phys.Rev} {\bf94} (2016) no.4, 043526
%DOI: 10.1103/PhysRevD.94.043526
%e-Print: arXiv:1605.04871 [astro-ph.CO] | PDF

%% A model of nonlinear electrodynamics
% S.I. Kruglov (Toronto U.). Sep 30, 2014. 8 pp.
% Published in Annals Phys. 353 (2014) 299-306
% DOI: 10.1016/j.aop.2014.12.001
% e-Print: arXiv:1410.0351 [physics.gen-ph]
%

\end{document}